\documentclass[manuscript]{acmart}
\AtBeginDocument{%
  \providecommand\BibTeX{{%
    \normalfont B\kern-0.5em{\scshape i\kern-0.25em b}\kern-0.8em\TeX}}}

\setcopyright{acmcopyright}
\copyrightyear{2023}
\acmYear{2023}
\acmDOI{XXXXXXX.XXXXXXX}

\acmBooktitle{28th Annual Conference on Intelligent User Interfaces (IUI '23), Mar 27--31, 2023, Sydney, Australia} 
\acmPrice{15.00}
\acmISBN{978-1-4503-XXXX-X/18/06}

\usepackage{amsmath}

\DeclareMathOperator*{\argmin}{arg\,min}

\usepackage{subcaption}
\usepackage{graphicx}
\usepackage{float}

\usepackage{hhline}
\usepackage{multirow}

\newcommand{\yali}[1]{[{\color{blue} Yali: \it #1}]}

\begin{document}

\title{NeuralSI: Neural Design of Semantic Interaction for Interactive Deep Learning}

\author{Yali Bian}
\email{yali.bian@intel.com}
\orcid{0000-0002-3335-9007}
\authornotemark[1]
\affiliation{%
  \institution{Intel Labs}
  \city{Santa Clara}
  \state{California}
  \country{USA}
}

\author{Rebecca Faust}
\email{rfaust1@tulane.edu}
\affiliation{%
 \institution{Tulane University}
 \city{New Orleans}
 \state{Louisiana}
 \country{United States}}

\author{Chris North}
\email{north@vt.edu}
\affiliation{%
 \institution{Virginia Tech}
 \city{Blacksburg}
 \state{Virginia}
 \country{United States}}

\renewcommand{\shortauthors}{Bian et al.}

\begin{abstract}
An increasing number of studies have utilized interactive deep learning as the analytic model of visual analytics systems for complex sensemaking tasks.
In these systems, traditional interactive dimensionality reduction (DR) models are commonly utilized to build a bi-directional bridge between high-dimensional deep learning representations and low-dimensional visualizations.
While these systems better capture analysts' intents in the context of human-in-the-loop interactive deep learning, traditional DR cannot support several desired properties for visual analytics, including out-of-sample extensions, stability, and real-time inference.
To avoid this issue, we propose the neural design framework of semantic interaction for interactive deep learning.
In our framework, we replace the traditional DR with a neural projection network and append it to the deep learning model as the task-specific output layer.
Therefore, the analytic model (deep learning) and visualization method (interactive DR) form one integrated end-to-end trainable deep neural network.
In order to understand the performance of the neural design in comparison to the state-of-the-art, we systematically performed two complementary studies, a human-centered qualitative case study and an algorithm-centered simulation-based quantitative experiment.
The results of these studies indicate that the neural design can give semantic interaction systems substantial advantages while still keeping comparable inference ability compared to the state-of-the-art model.
\end{abstract}



\maketitle

\section{Introduction}
Visual analytics (VA)~\cite{cook2005illuminating} enables human-AI collaborative sensemaking through interactive visualizations.
The standard VA framework consists of three separate components: the data, the visualization, and the analytic model~\cite{Sacha_Knowledge_2014}.
The analytic model captures the user's intents based on their visual interactions and abstractions from the raw data.
The visualization component is an automatic model that generates visualizations from the data and patterns extracted by the analytic model and serves as the primary interface between the analyst and the analytic model.
This framework has been widely used to build complex decision-making and data exploration systems~\cite{4389006,sacha2016visual}.

\begin{figure}[t]
    \centering
    \includegraphics[width=\columnwidth]{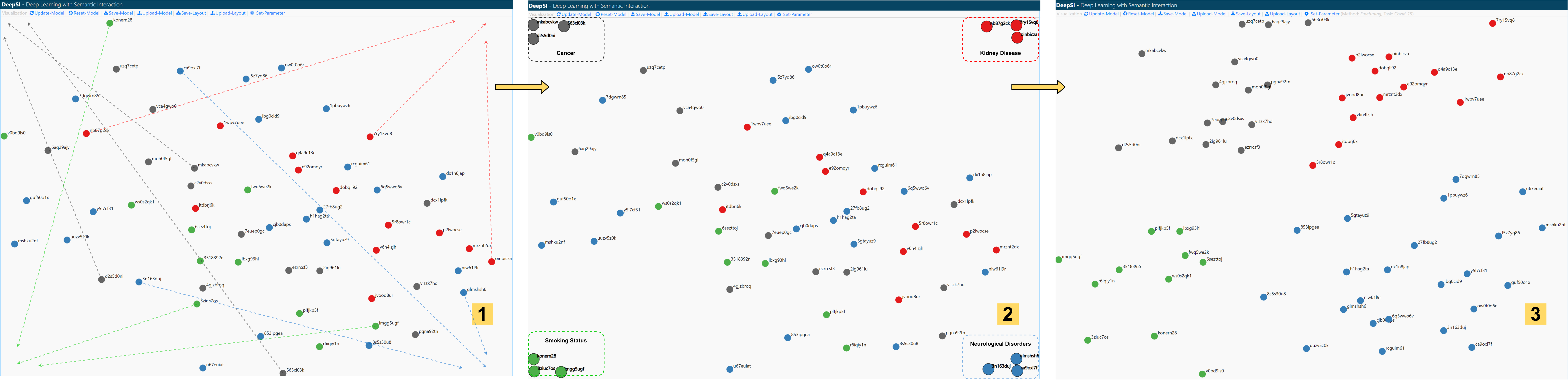}
    \caption{Screenshots during the analysis of COVID-19 research articles about four risk factors~(depicted in different colors) using DeepSI, an SI system with interactive DL: 
    (1) the initial layout of all articles projected from pretrained DL representations of the raw text data;
    (2) the analyst performs semantic interactions to provide visual feedback regarding articles about different risk factors; these interactions are then exploited to tune the underlying DL model via interactive DR;
    (3) the resulting projection updated by the tuned DL. (adapted from DeepSI~\cite{10.1145/3397481.3450670})}
    \label{fig:deepsi-covid}
\end{figure}

Semantic interaction (SI)~\cite{endert2012semantic} is a visual interaction methodology commonly applied to VA systems.
SI-enabled systems let the analyst directly manipulate interactive projections of data. 
Fig.~\ref{fig:deepsi-covid} shows an example semantic interaction where the analyst drags article points into four clusters to provide the visual feedback of grouping data points based on their preference (exploring articles about different risk factors of COVID-19).
The semantic meaning of these interactions indicates the relationships the analyst wishes to find within the data during the sensemaking process~\cite{pirolli_2005}.
It is the VA system's responsibility to capture the analyst's intent behind these interactions by learning a new projection layout~(Fig.~\ref{fig:deepsi-covid}-3).
Through these intuitive and natural interactions, the analyst remains within the cognitive zone~\cite{green2009building}, and the system thereby enhances the analyst's efficiency in performing analytic tasks~\cite{10.1145/3377325.3377516}. 

SI-enabled systems follow the standard VA framework in which
the analytic model utilizes a distance metric learning method to infer the intent behind the analyst's semantic interactions while
the visualization component, an interactive dimensionality reduction~(DR)~\cite{sacha2016visual}, acts as a bridge between the high-dimensional space updated by the metric learning method and the low-dimensional visual space manipulated by the analyst.
To determine the analyst's precise intent, increasingly powerful interactive DR models~\cite{10.1145/3377325.3377516} have been proposed, from linear models~\cite{Leman:2013it,6634115,House:2015hs} to non-linear models~\cite{10.1145/3311790.3396646,7534876}.

Deep learning~(DL)~\cite{LeCun:2015dt} is the state-of-the-art representation learning method~\cite{10.1109/tpami.2013.50} that automatically extracts abstract and useful hierarchical representations from raw data.
The capability of learning nonlinear patterns has boosted the integration of interactive DL into VA systems as the analytic model for complex sensemaking tasks~\cite{10.1109/tvcg.2019.2934595,10.1109/tvcg.2018.2865027}.
Previous research has shown that SI systems with interactive DL are better at inferring analysts' precise intents during sensemaking~\cite{bian2019evaluating,10.1145/3397481.3450670}.
These SI systems with interactive DL follow the standard VA framework, where the analytic model (DL model) and the visualization method (interactive DR) are two separate components.
During the human-DL co-learning process, these two components are executed individually in sequence: 
first, interactive DL generates user- and task-specific representations to capture analysts' precise intents;
then the interactive DR transforms the generated DL representations into a new visual layout totally from scratch.

Positively, SI systems use interactive DL to better capture analysts' intents~\cite{bian2019evaluating,10.1145/3397481.3450670}.
However, the high dimensionality and dynamic update properties of DL representations in the interactive co-learning process bring a new challenge. 
A small update to the DL model might cause a large, perceptually inconsistent, and inefficient change to the visualization, such as when applying non-deterministic dimension reduction like t-SNE~\cite{maaten2008visualizing}.
The usage of conventional DR models triggers potential incompatibilities that hinder desired properties such as out-of-sample extensions, stability, and real-time inference~\cite{Hinton504,hadsell2006dimensionality,espadoto2020deep}. 

This work aims to address this challenge by proposing a neural design framework for SI systems with interactive DL.
Specifically, our framework, NeuralSI, replaces the conventional DR model with a neural projection network. 
NeuralSI combines both the analytic model (interactive DL) and visualization method (interactive DR) into one integrated end-to-end trainable deep neural network.
The integrated DL model has two parts: 
(1) a pretrained DL as the backbone, which serves as the analytic model to infer users' concepts and update representation vectors;
and (2) a small neural network projection head appended to the backbone, which functions as the visualization method to communicate with analysts.
Based on the framework, we implement the SI system prototype for visual text analytic tasks. 
In the prototype, the backbone is a pretrained BERT model, a state-of-the-art DL model for NLP tasks.
The neural projection head is carefully designed based on three aspects: projection head architecture, projection parameter initialization, and loss functions.  
\yali{Updated here.}

To examine how well the NeuralSI prototype addresses the goals, we
measured its performance during the human-DL co-learning process in two respects: inference accuracy and time efficiency.
We compare it with the well-evaluated and state-of-the-art SI system with interactive DL, DeepSI~\cite{10.1145/3397481.3450670}, in two complementary experiments:
A human-centered qualitative case study about COVID-19 academic articles; 
and an algorithm-centered simulation-based quantitative analysis of three commonly used text corpora: Standford Sentiment Treebank~(SST), Vispubdata, and 20 Newsgroups. 
The results show that NeuralSI achieves comparable (slightly lower) inference accuracy while outperforming the state-of-the-art alternative in time efficiency.

Specifically, we claim the following contributions: 
\begin{enumerate}
    \item Identification of a neural design framework for SI systems with interactive deep learning, $\text{NeuralSI}$, that integrates the analytic model and visualization method into one end-to-end trainable neural network;
    \item  Two complementary studies, a user-centered qualitative case study and an algorithm-centered simulation-based quantitative experiment, that measure the inference accuracy and time efficiency of our method and reveal improvements.
\end{enumerate}


\section{Related Work}
\label{sec:related-works}
Three types of research support our design of the NeuralSI framework: 
VA systems with semantic interaction,
deep learning for visual analytics, 
and multidimensional projection with neural networks.

\begin{figure}[htpb]
    \centering
    \includegraphics[width=\columnwidth]{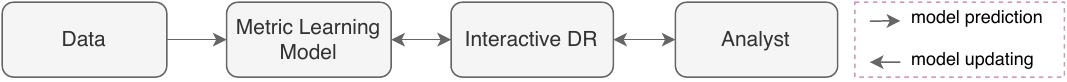}
    \caption{SI pipeline that follows the standard VA framework (adapted from \cite{sacha2016visual,endert2012semantic}).
    The interactive DR component serves as the visualization method responsible for capturing the intents behind analysts' interactions and updating the projection in response to the latest data relevance updated by the metric learning method.
    The metric learning method~\cite{10.1109/VAST.2012.6400486} is the analytic model responsible for inferring the intents from interactive DR by recalculating the data relevance and providing the updated data relevance as feedback.
    }
    \label{fig:si-pipeline}
\end{figure}

\subsection{VA Systems with SI}
\label{sec:si}
SI is commonly utilized to enhance VA systems for a variety of sensemaking tasks, including visual text analytics~\cite{endert2012semantic}, multivariate data analysis~\cite{self2015designing}, and visual concept explorations~\cite{bian2019deepva}.
Existing SI systems follow the standard VA framework, including the data, analytic model, and visualization components. 
As shown in Fig.~\ref{fig:si-pipeline}, the analytic model utilizes a distance metric learning method to infer the intent behind the analyst's interactions.
The visualization component is an interactive dimensionality reduction~(DR) method~\cite{sacha2016visual} which supports the bidirectional transformation between the high-dimensional feature space and the low-dimensional visual space~\cite{self2015designing,9623322}.

VA frameworks V2PI~\cite{Leman:2013it} and BaVA~\cite{House:2015hs} adapted linear DR models, including principal component analysis (PCA)~\cite{WOLD198737} and weighted multidimensional scaling (WMDS)~\cite{schiffman1981introduction}, to the bidirectional SI pipeline.
To support more complex tasks and interactions, more powerful but complicated nonlinear DR algorithms, such as t-SNE~\cite{maaten2008visualizing} and UMAP~\cite{mcinnes2018umap}, have been adapted in VA systems with SI~\cite{10.1145/3311790.3396646}.
To further improve the inference ability, Sharkzor~\cite{DBLP:journals/corr/abs-1802-05316} and DeepSI~\cite{10.1145/3397481.3450670} used DL as the analytic model in SI systems.
Different from existing works, we propose a new neural design framework of SI where both the analytic model and the visualization method are combined into one neural network.

\subsection{Deep Learning for Visual Analytics}
In contrast to conventional machine learning techniques, deep learning can automatically extract abstract and useful representations from raw data.
An increasing number of studies have explored the integration of deep learning and visual analytics in different fields. 
Research on the combination of VA and DL can be categorized into two areas: VA for DL and DL for VA.
For a more detailed illustration of VA for DL, we refer readers to the exhaustive surveys~\cite{hohman2018visual,8402187}.
This section only highlights previous relevant work on DL for VA: supporting visual analytics systems with DL for complex sensemaking tasks.

Hsueh-Chien et al. ~\cite{8265023} have used the CNN model~\cite{krizhevsky2012imagenet} to assist users in volume visualization designing by facilitating user interaction with high-dimensional DL features.
In RetainVis~\cite{10.1109/tvcg.2018.2865027}, an interactive and interpretable RNN model~\cite{hochreiter1997long} was designed for electronic medical records analysis and patient risk predictions.
Hyesook et al.~\cite{app11135853} have proposed a visualization system to analyze deep learning models with air pollution data according to the characteristics of spatiotemporal data.
Deep motif dashboard~\cite{lanchantin2017deep} provides a series of visualization strategies to sequence patterns extracted from three DL models (convolutional, recurrent, and convolutional-recurrent networks) to identify meaningful DNA sequence signals.
Gehrmann et al.~\cite{10.1109/tvcg.2019.2934595} have proposed a framework of collaborative semantic inference that enables visual collaboration between humans and DL algorithms. 

The existing works focus on integrating DL techniques as the analytic model in the standard VA framework for complex sensemaking tasks.
NeuralSI also uses the DL model as the analytic model, but rather than using a separate visualization component; it combines the analytic model and the visualization method into one integrated deep neural network. 

\subsection{Data Projection with Neural Networks}
\label{sec:mdp}
Multidimensional projection (MDP)~\cite{10.1109/tvcg.2018.2846735} is a commonly used data visualization technique for exploratory analysis of high-dimensional data. 
MDP transforms multidimensional data into scatter plots to reflect the data patterns in the original high-dimensional space.
A large amount of MDP techniques (such as PCA~\cite{WOLD198737}, MDS~\cite{schiffman1981introduction}, and t-SNE~\cite{maaten2008visualizing}) have been integrated into VA systems for different data exploration tasks, such as generating maps and exploring dimensions~\cite{10.1109/tvcg.2018.2846735,sacha2016visual}.
These traditional MDP methods have drawbacks~\cite{van2009dimensionality} in several perspectives including inference speed (computational time), out-of-sample~\cite{10.5555/2981345.2981368}, and consistency (smooth)~\cite{hadsell2006dimensionality}.

During the exploratory data analysis, analysts frequently update the visualization by loading new data and executing the DR method. 
These drawbacks can pose significant challenges in real-world VA systems.
For example, traditional MDPs (MDS and t-SNE, etc.) usually have a quadratic runtime~\cite{espadoto2020deep}.
In addition, the DR model has to recalculate the projection for the whole dataset when out-of-sample examples are loaded into the VA system.
It is infeasible to apply traditional MDPs to VA systems with large datasets that need real-time explorations. 
Furthermore, the updated visual projection can be dramatically different from the original data layout after the recalculation. 
The inconsistency between visual projections across different iterations can distract users from the primary analytic task. 

A series of neural networks including autoencoder~\cite{Hinton504}, DrLIM~\cite{hadsell2006dimensionality}, and neural network projections~(NNP)~\cite{espadoto2020deep}, have been proposed as replacements for traditional dimension reduction models.
These neural versions of visualization show substantial advantages for VA systems, such as out-of-sample extensions, stability, and inference speed~\cite{espadoto2020deep,10.1145/3311790.3396646,hadsell2006dimensionality}.
First, NNP has linear complexity and high scalability.
This enables NNPs to handle big data with high-dimensional representations effectively.
In addition, NNPs project data in a stable and consistent fashion across multiple iterations, which supports increased usability in the primary analytic task. 
As parametric models, NNPs need to be carefully initialized before usage.
Parameters inside NNPs can be initialized by training the projection network with the projections generated by traditional dimension reduction
methods~\cite{espadoto2020deep}, such as t-SNE~\cite{maaten2008visualizing}.
Another scheme for  initialization is self-supervised learning, such as the use of Autoencoders~\cite{Hinton504,kingma2013auto}.

The neural MDP has been used as the visualization method in VA systems for these desired properties.
For example, Zexplorer~\cite{10.1145/3311790.3396646} used a pretrained NNP as the interactive DR in SI systems.
Like Zexplorer, NeuralSI also uses an NNP as the interactive DR component.
However, NeuralSI appends the neural network (used as the visualization method) to the analytic model (DL) to form an integrated deep neural network.

\begin{figure*}[htpb]
    \centering
    \includegraphics[width=\textwidth]{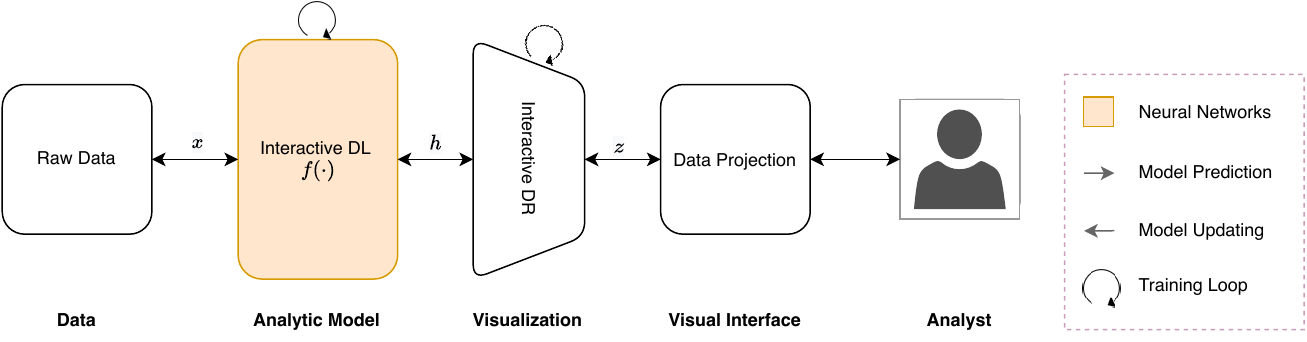}
    \caption{$\text{DeepSI}$ framework: embedding DL within the standard SI pipeline as the analytic model (adapted from~\cite{Sacha_Knowledge_2014,10.1145/3397481.3450670}).
    The interactive DR serves as the visualization component. 
    The usage of traditional DR models makes the visualization component separate from the analytic model. 
    During the co-learning process, DL and DR are in two different training loops and are required to compute separately.
    }
    \label{fig:deepsi}
\end{figure*}

\section{Background}
\label{sec:background}
In order to frame the discussion of our NeuralSI framework, this section briefly describes DeepSI, the state-of-the-art SI model that uses DL as the underlying analytic model.
As shown in Fig.~\ref{fig:deepsi}, the DeepSI pipeline follows the standard VA framework, which contains three components: data, DL as the analytic model, and interactive DR (specifically MDS) as the visualization method~\cite{Sacha_Knowledge_2014}.
DL is the analytic model responsible for generating user- and task-specific representations to capture analysts' precise intents.
The interactive DR (MDS) is responsible for updating the visual interface (data projection) based on the model patterns extracted by DL.
As the traditional DR, MDS is a nonparametric model and can directly pass the semantics of users' interactions to DL training without information loss.
This enables DeepSI a state-of-the-art performance. 
We illustrate the detailed human-DL co-learning process of using DeepSI as follows.

In the forward model-prediction direction, the analytic model (DL) generates new representation vectors for the raw data, $x$, through forward propagation with current DL parameters, ($\boldsymbol{w}_{\text{\tiny{backbone}}}$), given by the equation
\begin{equation}
    \label{eq:deepsi-forward}
    \boldsymbol h = \text{f}(x, w_{\text{\tiny{backbone}}})
\end{equation}
where $h$ is the output of the DL model $f(\cdot)$.
Next, the interactive DR (MDS) projects the high-dimensional DL representations $h$ to the 2D visual interface through the following equation (N is the total number of samples):
\begin{equation}
    \label{eq:deepsi-mds}
    \boldsymbol z = \argmin_{z}{\sum_{i<j\leq N}^{} \Big(||z_i - z_j|| - || h_i - h_j||\Big)^2}
\end{equation}
Importantly, for the updated $\boldsymbol{h}$ in each iteration, MDS needs to train the new low-dimensional embeddings $\boldsymbol{z}$ from scratch.
Based on the updates of $\boldsymbol{h}$, which represent the new out-of-sample data, a new MDS should be trained from scratch to learn $z$ in each loop.

In the backward model-updating direction, the analyst provides visual feedback by repositioning $n$ data points within the projection. 
Then, the interactive DR captures the human-defined similarities between $n$ moved data points, $dist(z_i, z_j)$, and uses them to steer the DL model parameters, $\boldsymbol{w}_{\text{\tiny{backbone}}}$, to generate better high-dimensional representations, $h$, such that the similarity of the representations reflects the proximity of the points in the modified projection, as follows: 
\begin{equation}
    \label{eq:deepsi-backward}
   w_{\text{\tiny{backbone}}} = \argmin_{w_{\text{\tiny{backbone}}}}{\sum_{{\tiny i<j\leq n}}^{} \Big(dist(z_i, z_j) - ||\text{f}(x_i)-\text{f}(x_j)) || \Big)^2}
\end{equation}
The optimization objective is to tune the DL weights, $\boldsymbol{w}_{\text{\tiny{backbone}}}$, to minimize the difference between low-dimensional and high-dimensional distances of $n$ moved data points through backpropagation. 
After the backpropagation, the updated $\boldsymbol{w}_{\text{\tiny{backbone}}}$ is used in the forward propagation to calculate the new representations, $\boldsymbol{h}$, shown in (\ref{eq:deepsi-forward}).

In DeepSI, the analytic model and visualization method are two separate components.
During the human-DL co-learning process, these two components are executed individually in sequence:
first, the interactive DL generates user- and task-specific representations to
capture analysts' precise intents;
then the traditional DR method, MDS, has to transform the generated DL representations to a new visual layout totally from scratch.
These two separate and sequential training loops cause the system's loss of the desired properties for visual analytics, including out-of-sample extensions, stability, and real-time inference~\cite{espadoto2020deep,10.1145/3311790.3396646,hadsell2006dimensionality} (discussed in section~\ref{sec:mdp}).
To avoid these drawbacks, we propose $\text{NeuralSI}$. 
Because NeuralSI uses neural networks for both the visualization method and the analytic model, it forms one integrated end-to-end trainable deep neural network.

\begin{figure*}[htpb]
\centerline{\includegraphics[width=\textwidth]{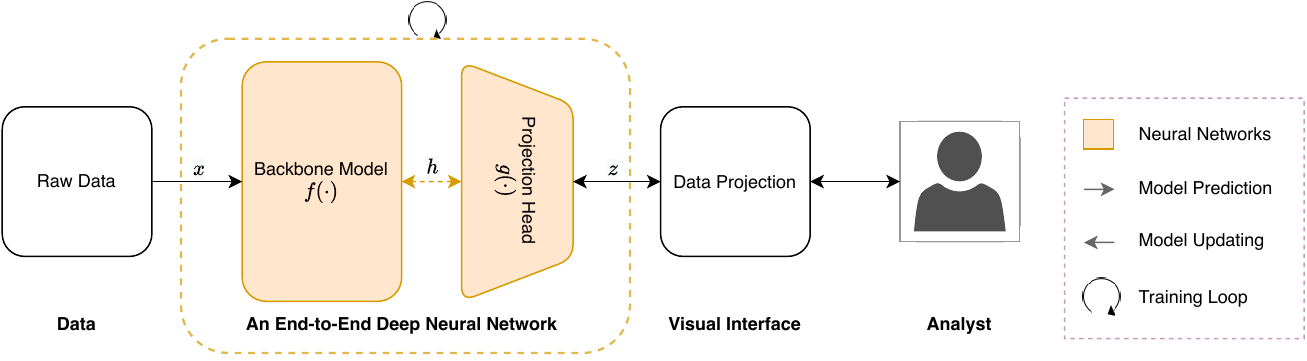}}
\caption{The NeuralSI framework uses a deep neural network as the analytic model and a relatively shallow neural network projection as the visualization method.
These two components (the analytic model and visualization) form one integrated end-to-end trainable neural network responsible for concept learning and visualization generation. 
The two components are in the same training loop and compute together through backpropagation.
}
\label{fig:neuralsi}
\end{figure*}

\section{Neural Design of SI}
\label{sec:neuralsi}
In this section, we first describe the NeuralSI framework where DL and DR are in the same training loop. 
We then represent the implementation details of the NeuralSI prototype, including the backbone model settings and the three design aspects of the projection head.

\subsection{The NeuralSI Framework}
\label{sec:framework}
We propose NeuralSI, a neural design VA framework of semantic interaction which uses an end-to-end trainable deep neural network.
The integrated deep neural network is responsible for both the analytic model and visualization method.
During the human-DL co-learning process, these two parts execute in the same training loop as an integrated unit to capture users' intents and generate new visualizations.
Therefore, NeuralSI avoids the limitations of existing systems and supports the desired properties, including out-of-sample extensions, stability, and real-time inference. 
Please refer to section~\ref{sec:mdp} for a more detailed illustration of these advantages of neural network projections over traditional data projection methods.

Inspired by SimCLR~\cite{chen2020simple}, we describe the integrated neural network in two parts based on their functionalities:
\begin{itemize}
    \item \textit{A backbone model} $f(\cdot)$ functioning as a deep neural network for extracting representation vectors, $h$, from input data examples, $x$, for downstream tasks. 
    In this model, $w_{\text{\tiny{backbone}}}$ represents the internal parameters and contains a large number of variables.
    The framework allows for various options in the network architecture without any constraints.
    The backbone model serves as the analytic model that infers users' intents, discovers data patterns, and presents them as visual feedback. 
    \item \textit{A projection head} $g(\cdot)$ which is a small neural network projection appended to the backbone model.
    The projection head maps the high-dimensional representations, $h$, to the visual projection, $z$, so that the user can gain insights from the visual layout. 
    The projection head is the visualization method for visual projection generation.
    $w_{\text{\tiny{projection}}}$ represents the internal parameters of the projection head.
    From the DL perspective, the projection head is the task-specific output layer for supervised DR~\cite{2006.05865}.
\end{itemize}

Fig.~\ref{fig:neuralsi} illustrates the detailed usage of NeuralSI through the human-DL co-learning process.
Before the co-learning process, the two parts (the backbone model and the projection head) of the deep neural network are initialized separately.
The parameters of the backbone model, $w_{\text{\tiny{backbone}}}$, are trained with a massive number of data to learn general representations that can then be easily adapted to downstream tasks.
The initialization of the projection head paramters, $w_{\text{\tiny{projection}}}$, is described in section~\ref{sec:init}.

In the forward model-prediction direction, the end-to-end neural network directly generates the visual layout $z$ (2D data projection) from $x$, with the DL parameters~($\boldsymbol{w}_{\text{\tiny{backbone}}}$ and $\boldsymbol{w}_{\text{\tiny{projection}}}$):
\begin{equation}
    \label{eq:neuralsi-forward}
    \boldsymbol z = \text{g}(\text{f}(x, w_{\text{\tiny{backbone}}}), w_{\text{\tiny{projection}}})
\end{equation}
As shown in (\ref{eq:neuralsi-forward}), the high-dimensional DL representations $h$, calculated by $\text{f}(x)$, are directly passed to the projection head, and are then transformed to the 2D spatialization by $\text{g}(h)$.

In the backward model-updating direction, the analyst modifies the visual layout by repositioning samples to express the preferred similarities.
Then, the loss function $L$ (described in section~\ref{sec:loss}) uses the moved data points, $z_{i}$, to steer the parameters of the end-to-end neural network, $g(f(x))$.
The process is illustrated in the following equation: 
\begin{equation}
    \label{eq:neuralsi-backward}
   w = \argmin_{w}{\mathcal{L}(g(f(x)), z)}
\end{equation}
The optimization objective is to fine-tune DL parameters (both $\boldsymbol{w}_{\text{\tiny{backbone}}}$ and $\boldsymbol{w}_{\text{\tiny{projection}}}$) to minimize the difference between predicted $\hat z$ and the user-updated $z$ variables.
All internal parameters of the DL model~(i.e., $\boldsymbol{w}_{\text{\tiny{backbone}}}$ for $\text{f}$, and  $\boldsymbol{w}_{\text{\tiny{projection}}}$ for $\text{g}$) are updated in order, from the projection head to the backbone, by a gradient descent optimization algorithm. 
After the backpropagation, the updated parameters are used in the forward propagation to calculate new projections ($\boldsymbol{z}$ in Equation~\ref{eq:neuralsi-forward}).

\subsection{Prototyping Detail}

To measure the performance of NeuralSI, we applied visual text analytic tasks to the system.
There are two reasons for this. 
First, VA systems with SI are primarily used for visual text analytics tasks, such as intelligence analysis~\cite{10.1109/vast.2011.6102449} and research publication exploration~\cite{10.1145/3311790.3396646}.
In addition, the state-of-the-art SI system, DeepSI, has been evaluated with visual text analytic tasks, making it a suitable benchmark for comparison purposes.

To implement the NeuralSI system in visual text analytic tasks, we used the pretrained BERT~\cite{devlin2018bert}, a state-of-the-art DL model for NLP tasks, as the default backbone of the framework.
In addition, to compare with the default MDS model used in DeepSI, an equivalent projection head was proposed for NeuralSI in three design aspects:
projection head architecture, projection parameter initialization, and loss function. 
Further, we used the default Adam optimizer~\cite{kingma2014adam} to optimize the model parameters inside the NeuralSI ($w_{\text{\tiny{backbone}}}$ and $w_{\text{\tiny{projection}}}$). 
We used the suggested learning rate~($3e^{-5}$) for the fine-tuning of the BERT models ~\cite{devlin2018bert} in our experiments. 
The following describes the settings of the backbone model and projection head in more detail.

\subsubsection{Backbone model}
The pretrained BERT model is adapted from the publicly available Python library, Transformers~\cite{Wolf2019HuggingFacesTS}.
Transformers provides two sizes of pretrained BERT models: $\text{BERT}_{\text{BASE}}$, and $\text{BERT}_{\text{LARGE}}$. 
We used the small BERT model~($BERT_{BASE}$)~(bert-base-uncased, 12-layers, 768-hidden, 12-heads, 110M parameters) for its stability on small datasets.
For a document containing a list of tokens, $\text{BERT}_{\text{BASE}}$ can convert each of the tokens into a 768-dimensional vector. 
To generate fixed-length encoding vectors from documents of different lengths, we appended a MEAN pooling layer to the last transformer layer of the BERT model.

\subsubsection{Projection Head}
Here we explore three major design aspects to make the projection head equivalent to MDS used in DeepSI: the projection head architecture, projection parameter initialization, and loss function.

\textbf{Projection Head Architecture: } 
\label{sec:struct}
The projection head, $g(\cdot)$, is the output layer of NeuralSI particularly designed for the downstream task~\cite{wu2018unsupervised,bachman2019learning}---interactive DR~\cite{8440814,9623322,self2015designing}.
As described in section~\ref{sec:mdp}, a variety of neural network architectures have been proposed for DR~\cite{Hinton504,hadsell2006dimensionality,espadoto2020deep}.
To compare with the default MDS model used in DeepSI, we constrained the architecture of the projection head to the basic one-layer linear projection.
As shown in (\ref{eq:linear-projection}), linear projection maps the high-dimensional representation, $h$, to the 2D projection, $z$:
\begin{equation}
    \label{eq:linear-projection}
    z = g(h) = w_{\text{\tiny{projection}}}*h
\end{equation}
Through linear transformation, a consistent data structure for the high-dimensional space is preserved in the 2D visualization.
Please note that normalization is needed after the transformation to fit the projected outputs to the limited visual space.


\textbf{Projection Parameter Initialization: }
\label{sec:init}
Unlike the nonparametric DR methods used in DeepSI, the nueral projection head of NeuralSI contains trainable paramters $w_{\text{\tiny{projection}}}$.  
Projection parameter initialization is critical to the overall performance of NeuralSI systems.
First, before the human-DL co-learning process, an effective initialization scheme can initially generate a moderate projection layout for data explorations. 
In addition, the properly initialized weights of the projection head also enhance the performance of the subsequent co-learning process.

Neural network projections use two common initialization schemes: 
(1) initialization from the projections generated by traditional MDPs (discussed in section~\ref{sec:mdp}), such as T-SNE~\cite{maaten2008visualizing}, 
and (2) self-supervised learning, such as the use of Autoencoders~\cite{Hinton504}.
These two initialization schemes have equal performance, as discussed in the Generalized Autoencoder~(GAE) framework~\cite{6910027}.
To simplify the comparison with DeepSI, we initialized the projection head by training its parameters with the MDS projection results.
We call this method MDS-based initialization.

\textbf{Loss Functions: }
\label{sec:loss}
NeuralSI systems utilize users' visual interactions on the data projections as the instructions to tune the integrated deep neural network $g(f(x))$.
Through repositioning projected data points ($z$) on the visual space, analysts intuitively express their task-specific preferences for the projection.
When regarding the SI training as a deep metric learning task, the existing loss functions that could be employed in NeuralSI include contrastive loss, triplet loss~\cite{10.1007/978-3-319-24261-3_7} and N-pair loss~\cite{sohn2016improved}.
In the deep metric learning task, the similarity information comes from the repositioned data points on the visual layout.
These repositioned data points can guide models to generate projections with the desired ``similarity $\approx$ proximity" property.
For purposes of simplicity and comparison, we use the most commonly used loss function: contrastive loss~\cite{chen2020simple,10.3390/sym11091066}. 
It is also the primary loss function used in existing SI systems~\cite{10.1109/VAST.2012.6400486,self2018observation,10.1145/3397481.3450670}, and can be expressed as:
\begin{equation}
    \label{eq:contrastive-loss}
    \mathcal{L}_{contrastive} = \sum_{i<j\leq n}^{} \Big(||z_i - z_j|| - ||\hat z_i - \hat z_j||\Big)^2
\end{equation}
where $z_{i}$ are the positions of all moved data points, and $\hat z_i$ are the positions predicted by the projection $g(f(x))$.
Contrastive loss enables NeuralSI only to need neighborhood relationships between the moved samples.


\section{Experiments}
\label{sec:experiments}

To examine how well NeuralSI addresses the goals discussed in section~\ref{sec:framework},
we measured its performance in two respects during the human-DL co-learning process: 
\begin{itemize}
    \item RQ1: How does the inference accuracy of NeuralSI compare with the state-of-the-art SI system DeepSI?
    \item RQ2: How does the time efficiency of NeuralSI in response to analysts' interactions compare with DeepSI? 
\end{itemize}

For RQ1, we used two complementary evaluation methods to measure the inference accuracy: the case study in section~\ref{sec:case-study} is the human-centered qualitative analysis, and the simulation-based evaluation method in section~\ref{sec:simulation} is the algorithm-centered quantitative analysis.
For RQ2, we measured and compared the running time of SI systems in performing tasks with datasets of different sizes in section~\ref{sec:running-time}.

\subsection{Case Study: COVID Paper Exploration}
\label{sec:case-study}
Recently, COVID-19~\cite{doi:10.1177/0020764020915212} has become a global pandemic. 
It is essential that medical researchers quickly find relevant documents about a specific research question, given the extensive coronavirus literature.
We used an analysis task on academic articles related to COVID-19 in this case study to examine our proposed NeuralSI, compared with the state-of-the-art SI system, DeepSI in terms of inference accuracy (RQ1).

\subsubsection{Dataset and Task}
\label{sec:covid-19}
The COVID-19 Open Research Dataset~(CORD-19)~\footnote{https://www.kaggle.com/allen-institute-for-ai/CORD-19-research-challenge} contains a collection of more than $400,000$ academic articles about COVID-19. 
CORD-19 also proposes a series of tasks in the form of important research questions about the coronavirus. 
One of the research tasks focuses on identifying COVID-19 risk factors~\footnote{https://www.kaggle.com/allen-institute-for-ai/CORD-19-research-challenge/tasks?taskId=558}. 
In this case study, we selected a task that requires identifying articles related to specific risk factors for COVID-19. 
We asked an expert to choose as many research papers as possible about risk factors from CORD-19.
We focused on four main risk factors in this task: 
cancer~(15 articles), 
chronic kidney disease~(13 articles), 
neurological disorders~(23 articles), 
and smoking status~(11 articles). 
We used these four risk factors as the ground truth for the test task and loaded all these 62 articles into these two SI systems.
Therefore, the test task was to organize these 62 articles into four clusters with the SI tool such that each cluster represented articles of a specific risk factor.

This particular ground truth is just one possible way an analyst might want to organize this group of documents. 
Our goal is to inspect how easily this particular set of expert knowledge can be injected using SI to help re-organize the documents accordingly.
To help judge the quality of the visual layout in organizing this particular ground truth, we color the dots according to the ground-truth risk factors in the visualization:
\textbf{cancer}~(black dot $\textcolor{black}{\bullet}$), 
\textbf{chronic kidney disease}~(red dot $\textcolor{red}{\bullet}$),
\textbf{neurological disorders}~(blue dot  $\textcolor{blue}{\bullet}$),
\textbf{smoking status}~(green dot $\textcolor{green}{\bullet}$).
It should be noted that the underlying model was not provided with ground truth or color information. 
The ground truth is only injected via semantic interaction from the human in the form of partial groupings of only a few of the documents.

\subsubsection{Study Procedure}
Similar semantic interactions based on the ground truth are performed in both visual projections and applied to tune the underlying models to compare the projection layouts updated by two SI systems.
Fig.~\ref{fig:deepsi-covid} and Fig.~\ref{fig:neuralsi-covid-1} show the process of interactions applied separately to DeepSI and NeuralSI prototypes.
In both figures, frame 1 shows the initial layout: Fig.~\ref{fig:deepsi-covid}-1 is projected by the MDS method, and Fig.~\ref{fig:neuralsi-covid-1}-1 is predicted by the neural projection head.
In both initial projection layouts, all the articles are combined.
This means that the pretrained BERT model cannot distinguish these articles by their related risk factors.
Furthermore, these two frames still have similar layouts while being updated by different DR methods.
The reason is that the neural projection head is initialized by learning an MDS projection result. 

Interactions were performed within these projections based on the ground truth to reflect the perceived connections between articles:
grouping three articles about cancer to the top-left region of the projection, indicated by the black arrows;
three articles about chronic kidney disease to the top-right region indicated by red arrows;
three articles about smoking status to the bottom-left part indicated by green arrows;
and three articles about neurological disorders to the bottom-right part indicated by blue arrows.

Frame 2~(Fig.~\ref{fig:deepsi-covid}-2 and Fig.~\ref{fig:neuralsi-covid-1}-2) shows the human spatializations which creates four clusters based on the ground truth.
We then click the update-model button, using the human spatializations to train the two models (DeepSI and NeuralSI).
After updating the models, the projections update to show how it incorporated the injected knowledge, shown in frame 3 (Fig.~\ref{fig:deepsi-covid}-3 and Fig.~\ref{fig:neuralsi-covid-1}-3). 
Subsequently, we assess the performance of each model based on how reasonable the layout is in comparison with the ground truth. 



\textbf{DeepSI spatialization:} 
The projection updated by DeepSI is shown in Fig.~\ref{fig:deepsi-covid}-3.
There are four clear clusters, and all articles are clearly grouped into the correct clusters. 
The top left cluster contains all the articles about cancer~($\textcolor{black}{\bullet}$), 
the top right cluster contains articles about kidney disease~($\textcolor{red}{\bullet}$),
the bottom left contains articles about smoking status~($\textcolor{green}{\bullet}$),
and the bottom right contains articles about neurological disorders~($\textcolor{blue}{\bullet}$).
This means the new projection generated by DeepSI is able to capture the semantic meanings behind users' interactions accurately. 

\begin{figure*}[htpb]
    \centering
    \includegraphics[width=\textwidth]{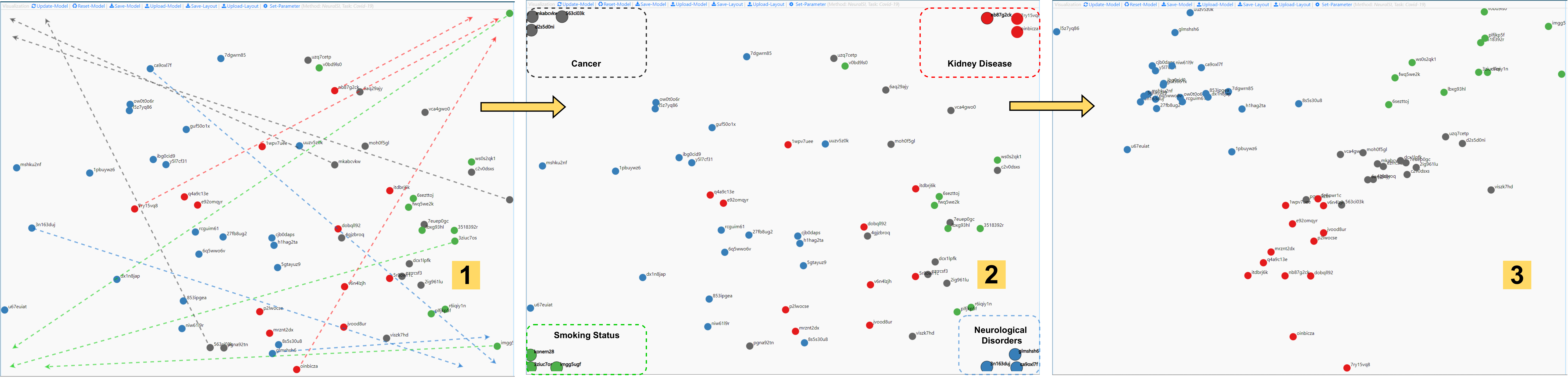}
    \caption{Screenshots during the case study using NeuralSI: 
    Frame 1 is the initial layout of all articles predicted by the integrated network;
    Frame 2 shows the similar interactions performed by the analyst in~Fig.~\ref{fig:deepsi-covid}-2. 
    Frame 3 shows the resulting projection updated by NeuralSI.}
    \label{fig:neuralsi-covid-1}
\end{figure*}

\textbf{NeuralSI spatialization:} 
With the same interactions as input, the updated NeuralSI shows a slightly different layout.
As shown in Fig.~\ref{fig:neuralsi-covid-1}-3, there are four clusters in the updated layout.
However, two clusters (kidney disease $\textcolor{red}{\bullet}$ and cancer $\textcolor{black}{\bullet}$) overlap slightly and are not clearly separated compared with Fig.~\ref{fig:deepsi-covid}-3.
To properly capture the user's semantic intent and differentiate these two clusters, continued interactions based on the ground truth are needed. 

\begin{figure*}[htpb]
    \centering
    \includegraphics[width=\textwidth]{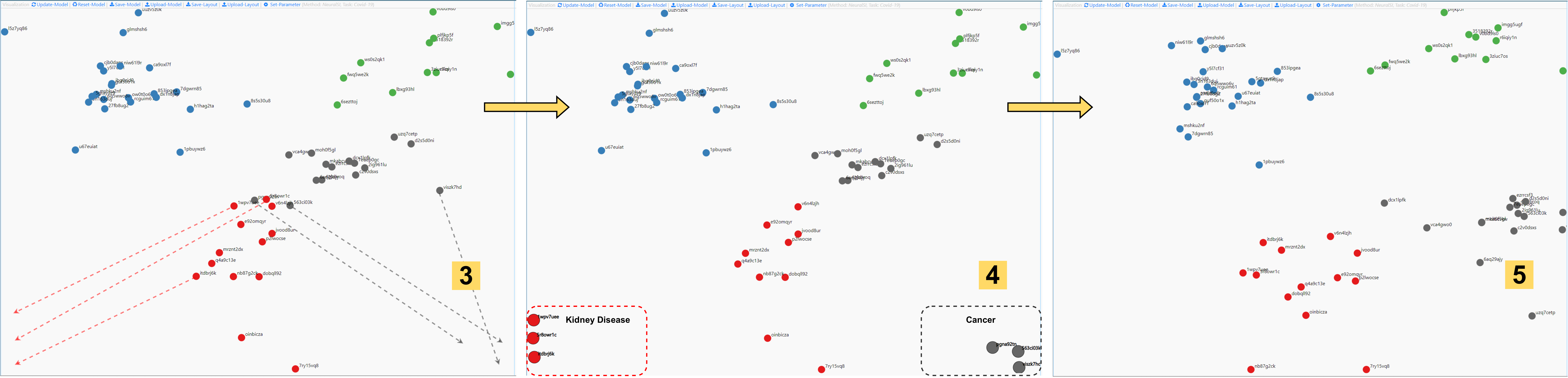}
    \caption{Continued interactions performed by the analyst in separating two clusters using NeauralSI: 
    Frame 3 is the projection made by the tuned NeuralSI in the first iteration,
    Frame 4 shows the continued interactions performed within the projection in the second iteration, 
    and Frame 5 shows the resulting projection updated by NeuralSI.
    }
    \label{fig:neuralsi-covid-2}
\end{figure*}

\textbf{Continued interactions applied to NeuralSI:} 
As shown in Fig.~\ref{fig:neuralsi-covid-2}-3, in the second iteration, three kidney disease articles~($\textcolor{red}{\bullet}$) are dragged to the bottom-left and three cancer articles~($\textcolor{black}{\bullet}$) are dragged to the bottom-right on the scatterplot.
Fig.~\ref{fig:neuralsi-covid-2}-4 shows the human-created clusters based on the ground truth to emphasize the separation of data points in these two clusters.
After the layout updates, articles from these two dragged clusters are well placed on two opposite sides of the visualization in Fig.~\ref{fig:neuralsi-covid-2}-5.
Taking articles about the other two risk factors~(about neurological disorders $\textcolor{blue}{\bullet}$ and smoking status $\textcolor{green}{\bullet}$) into consideration, four clusters are well formed and clearly separated.

\subsubsection{Qualitative Inference Accuracy Comparison:}
In terms of accuracy, DeepSI grouped articles correctly based on user-defined risk factors with interactions in one iteration. 
In contrast, NeuralSI did not provide a perfect projection with four clearly separated clusters in one iteration.
Two clusters overlap slightly, which requires more cognitive effort to identify the boundary between the groups.
Continued interactions in another exploration iteration are needed to separate these two.
With more interactions as input, more distinguishable clusters are formed: articles within the same cluster gather more close, and data points in different clusters are more separate from each other.
NeuralSI generates the data projection that makes the semantics of the ground truth knowledge provided by the contest organizers more recognizable.
Compared with DeepSI, slightly more interactions are required to enable NeuralSI to capture users' intents accurately.

\subsection{Simulation-Based Evaluation}
\label{sec:simulation}

\subsubsection{Experiment Design}
SI systems interactively learn projections provided by the analyst in an online learning process~\cite{hoi2021online}.
To perform quantitative comparisons of these two SI systems, we employ the simulation-based evaluation method in the experiments.
This algorithm-centered method replaces the analyst with a simulation component.
In each iteration, the simulated analyst first randomly picks three samples from each class of the dataset and applies semantic interactions on these picked samples to train the SI system.
It then conducts a quantitative measurement of the updated visual projection produced by the tuned SI system. 
Specifically, we used a kNN~(K-nearest-neighbour) classifier~\cite{Cover:2006:NNP:2263261.2267456} to measure the quality of predicted projection~\cite{10.1109/VAST.2012.6400486,bian2019evaluating}, which reflects the similarity relationships between samples in the low-dimensional spatialization.
In each training iteration, the simulated analyst outputs a current projection accuracy.
The projection accuracy over iterations reflects the learning curve~\cite{10.1162/153244304322972694} of the SI system and is used to make systematic comparisons.
Admittedly, compared with human-centered analysis, the simulation-based method can only measure the performance of SI systems in clustering tasks.
However, the quantitive comparisons enable this method to provide an objective analysis, complementary to the above case study.
For a more detailed illustration of the simulation-based evaluation, please refer to~\cite{bian2019evaluating,10.1145/3397481.3450670}.
For the simulation process, we performed 200 interaction iterations.

\subsubsection{Dataset and Task}
To address RQ1, we quantitively compare the inference accuracy between NeuralSI and DeepSI using the simulation-based evaluation method on three text analytic tasks:
Stanford sentiment treebank (SST) dataset~\cite{socher2013recursive} with two clusters (positive and negative), denoted as $T_{\text{sst}}$; the Vispubdata dataset~\cite{Isenberg:2017:VMC} with three clusters (scivis, infovis, and vast), denoted as $T_{\text{vis}}$; and 20 newsgroup dataset\footnote{http://qwone.com/~jason/20Newsgroups/} with four clusters (from the 'rec' category: autos, motorcycles, sport.baseball, sport.hockey), denoted as $T_{\text{news}}$.
Among these three target tasks, $T_{\text{sst}}$ and $T_{\text{news}}$ are closer to the pre-training tasks of BERT model,
while $T_{\text{vis}}$ is more domain-specific and especially difficult.


\begin{figure*}[htpb]
    \begin{subfigure}[tb]{0.325\textwidth}
        \includegraphics[width=\textwidth]{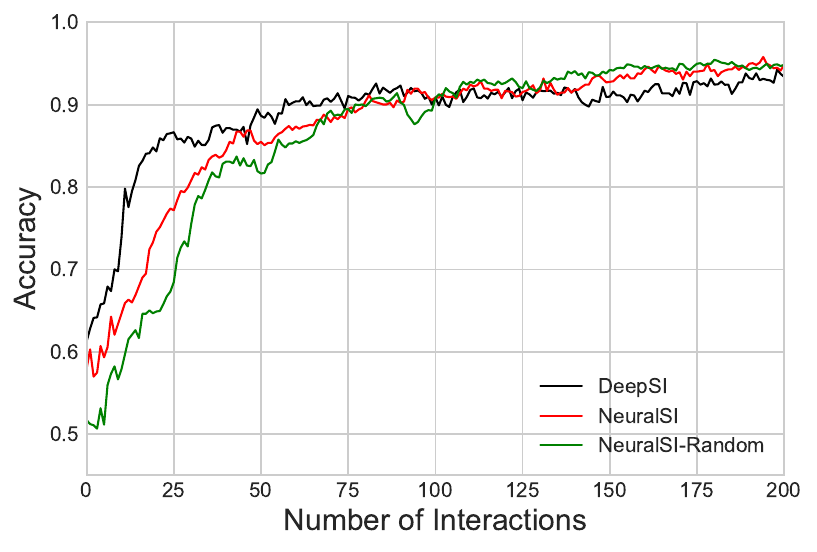}
        \caption{$T_{\text{sst}}$}
        \label{fig:task-loss-layout-sst}
    \end{subfigure}
     \begin{subfigure}[tb]{0.325\textwidth}
        \includegraphics[width=\textwidth]{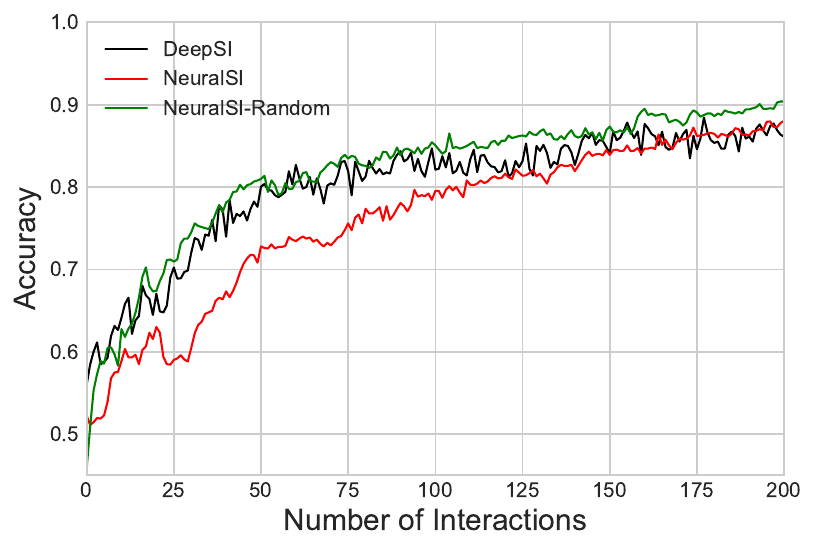}
        \caption{$T_{\text{vis}}$}
        \label{fig:task-loss-layout-vis}
    \end{subfigure}
     \begin{subfigure}[tb]{0.325\textwidth}
        \includegraphics[width=\textwidth]{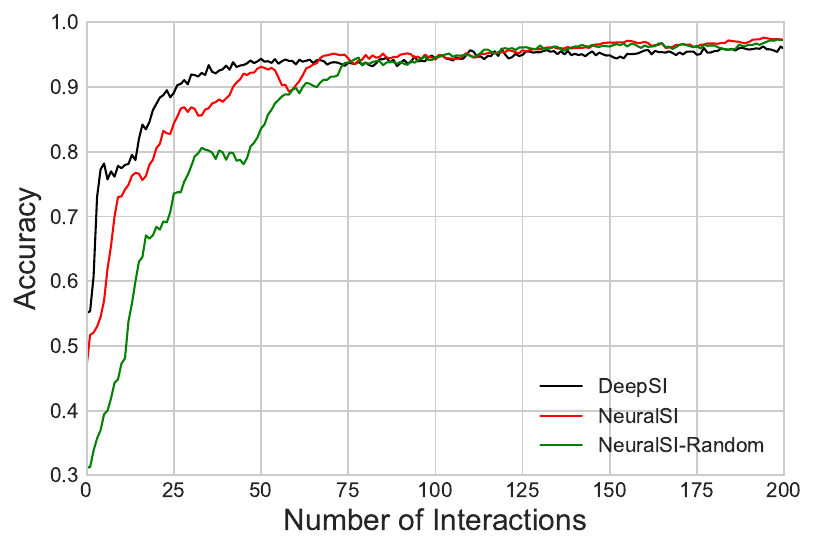}
        \caption{$T_{\text{news}}$}
        \label{fig:task-loss-layout-news}
    \end{subfigure}
    \caption{The accuracies of the $\text{DeepSI}$ and $\text{NeuralSI}$ models' updated projections over 200 iterations for the three tasks~($T_{\text{sst}}$, $T_{\text{vis}}$, and $T_{\text{news}}$) during the simulation-based experiment. NeuralSI is the default NeuralSI system where the project head is specified by MDS-based initialization. NeuralSI-Random is the NeuralSI system where the project head is defined by random initialization.}
    \label{fig:overall-results}
\end{figure*}


\subsubsection{Quantitative Inference Accuracy Comparison}
Fig.~\ref{fig:overall-results} shows the performance comparison between NeuralSI with DeepSI. 
Overall, DeepSI (black line) showed slightly better accuracy performance than NeuralSI (red line).
This is consistent with our findings in the case study~(Sec.~\ref{sec:case-study}), where DeepSI with a nonparametric DR gains a better accuracy than NearualSI systems with parametric projection heads. 
The result is evident because DeepSI uses the traditional nonparametric DR  as the visualization method.
There was no need to train extra parameters aside from those inside the backbone model.
With more interactions, the gap between DeepSI and NeuralSI decreases.
This is also consistent with the findings of few-shot learning, indicating that the nonparametric task-specific output layer is better at training DL models with fewer examples~\cite{10.1145/3386252,10.18653/v1/2020.coling-main.92}.

In addition, DeepSI only showed slightly better performance over NeuralSI in the two general-purpose tasks ($T_{\text{sst}}$ and $T_{\text{news}}$).
The gap between DeepSI and NeuralSI decreases with continued interaction, and after 75 interaction loops, NeuralSI even had slightly better performance until approximating to the peak accuracy in all three tasks.
In the domain-specific task, $T_{\text{vis}}$, DeepSI showed clearly better performance over NeuralSI,  
even after continued interactions based on the ground truth.
We conjectured that this performance inconsistency between the domain-specific task and two general-purpose tasks is triggered by the initialization method employed in the current NeuralSI system. 
The MDS-based initialization lets the neural projection head preserve the data patterns from the high-dimensional representations of the pretrained backbone.
If the task is too domain-specific, the initialized projection head might hinder the further fine-tuning of the end-to-end deep neural network.

To verify our hypothesis, we compared the default NeuralSI and DeepSI to NeuralSI with the randomly initialized projection head, called $\text{NeuralSI}_{\text{random}}$. 
Specifically, we used the robust initialization method proposed by Kaiming et al.~\cite{He_2015_ICCV}, known as He Initialization, for the projection head.
As shown in Fig.~\ref{fig:overall-results}, when the number of interactions is 0, $\text{NeuralSI}$ had better initial accuracies for all three tasks.
This indicates that MDS-based initialization allowed the projection head to generate better projections for data exploration before the co-learning process.
With enough interactions, the accuracy of each model became more similar as training interactions increased.
With the small number of interactions, $\text{NeuralSI}$ kept a significant advantage over $\text{NeuralSI}_{\text{random}}$ in two general purpose tasks ($T_{\text{sst}}$ and $T_{\text{news}}$).
In the domain-specific task, $T_{\text{vis}}$, 
$\text{NeuralSI}$ had lower accuracy gains than $\text{NeuralSI}_{\text{random}}$ when learning from training interactions.
Furthermore, $\text{NeuralSI}_{\text{random}}$  showed comparable accuracy performance to DeepSI. 
For general-purpose datasets and tasks, MDS-based initialization could increase the speed of the learning process without hindering the accuracy gains from training with more visual interactions.
However, if the analysis task is domain-specific, it hinders the training performance of NeuralSI systems.
This finding is consistent with the transfer learning and domain adaptation methodologies of reusing pretrained deep neural networks on new tasks~\cite{mou2016transferable}. 

\subsection{Time Efficiency Measurements}
\label{sec:running-time}

\subsubsection{Experiment Design.}
Semantic interaction systems, including NeuralSI and DeepSI, are intended for use in sensemaking tasks\cite{10.1109/tvcg.2019.2934595,10.1109/tvcg.2018.2865027,10.1145/3397481.3450670}.
In these tasks, analysts interactively explore and make sense of data sets, such as intelligence reports. 
During this exploratory analysis process, it is essential to make sure SI systems support near real-time exploratory interactions. 
Therefore, we evaluate and compare the time efficiency between NeuralSI and DeepSI in the three tasks ($T_{\text{sst}}$, $T_{\text{vis}}$ and $T_{\text{news}}$).

In real-world tasks, analysts will explore a reasonable number of data points that can be interacted with in the projection before applying the learned model to much larger datasets. 
Therefore, we measure the time efficiency in performing these interactive tasks with a reasonable data size between $100$ to $1000$. 
Specifically, we used data sizes of 100, 200, 500, and $1000$ data points.
We measured the average computation time of the SI system in response to each simulated interactive feedback, repeated 20 times.
As in section~\ref{sec:simulation}, the simulated feedback was performed on three randomly chosen samples from each class of the dataset.
All our experiments were conducted on a desktop computer with an Intel i9-9900k processor, 32G Ram, and one NVIDIA GeForce RTX 2080Ti GPU, running Windows 10 Pro.


\begin{table*}[t]
  \centering
  \renewcommand{\arraystretch}{1.2}
  \begin{tabular}{|p{1.5cm}|c|c|c|c|c|c|c|c|c|c|c|c|}
    \hline
    \multirow{2}{5cm}{\textbf{Modes}} & \multicolumn{4}{c|}{\textbf{$T_{\text{sst}}$}} & \multicolumn{4}{c|}{\textbf{$T_{\text{vis}}$}} & \multicolumn{4}{c|}{\textbf{$T_{\text{news}}$}} \\
    \cline{2-13}
    & \textbf{100} & \textbf{200} & \textbf{500} & \textbf{1000} & \textbf{100} & \textbf{200} & \textbf{500} & \textbf{1000} & \textbf{100} & \textbf{200} & \textbf{500}  & \textbf{1000}    \\
    \hline
    \textbf{DeepSI} & 2.10 & 4.14  & 34.97 & 152.19 & 2.63 & 4.52 & 30.32 & 139.37 & 2.48 & 4.26 & 23.65 & 92.91 \\ \hline
    \textbf{NeuralSI} & 0.33 & 0.54 & 1.25 & 2.56 & 1.76 & 3.00 & 6.84 & 13.44 & 1.77 & 3.04 & 6.92 & 13.57\\ \hline
  \end{tabular}
  \caption{Time consumptions (in seconds) of SI systems in reponse to three tasks with four different  dataset sizes.}
  \label{tab:time-efficiency}
\end{table*}

\subsubsection{Time Efficiency Comparison.}
\begin{figure}[htpb]
    \centering
    \includegraphics[width=0.75\columnwidth]{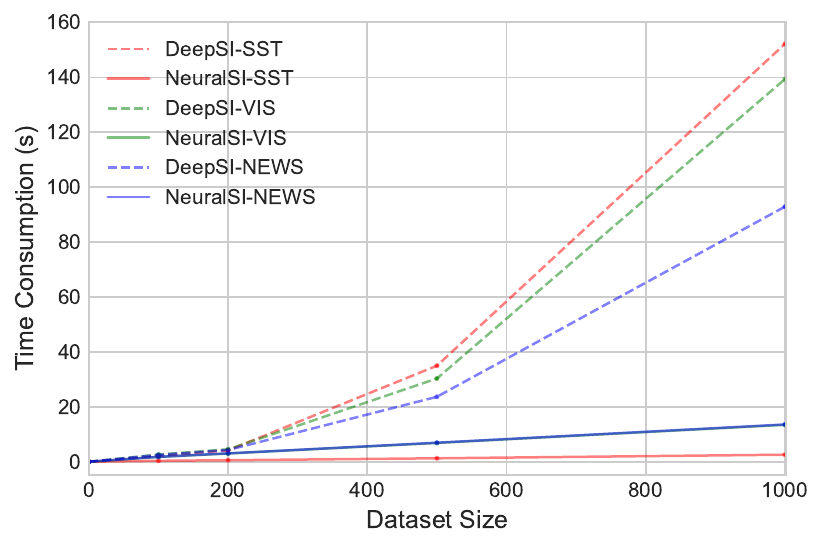}
    \caption{Time consumption of SI systems in response to three tasks with four different dataset sizes. The solid green line for NeuralSI-SST is occluded by the solid blue line of NeuralSI-NEWS because they have similar time consumption.
    }
    \label{fig:time-complexity}
\end{figure}

As shown in the table~\ref{tab:time-efficiency}, DeepSI took more time than NeuralSI to respond to interactive feedback and update the visual projections.
For example, NeuralSI responded in less than 14 seconds for all three tasks with $1000$ data points.
In particular, for datasets with short documents ($T_{sst}$), NeuralSI completes the whole iteration (the model updating and prediction process) in under 3 seconds.
However, it took at least 1.5 minutes for DeepSI to finish the visual feedback. 
In addition, Fig.~\ref{fig:time-complexity} shows that as data size increases, the time consumption of DeepSI increases slowly at first and then more rapidly which indicates a non-linear relationship between data size and time consumption.  
In contrast, the time consumption of NeuralSI shows a nearly linear relationship with data size that intercepts the origin.
It is worth mentioning that DeepSI spent more time on $T_\text{sst}$ with shorter documents.
This indicates that the DR component consumes a majority of the response time instead of the BERT backbone. 

NeuralSI has better scalability. It has better time efficiency, especially in bigger datasets than DeepSI.
In detail, NeuralSI has a time complexity of $O(n)$, where $n$ is the data size.
In contrast, the data projection (MDS) of DeepSI has a time complexity of $O(n^2)$, which consumes a majority of the response time.
In addition, NeuralSI uses the GPU to accelerate the whole model, while DeepSI uses the CPU to execute the traditional dimension reduction algorithm (MDS). 

\section{Discussion}

\subsection{Trade-off between Accuracy and Scalability}

NeuralSI requires a few more interactions from users to capture their intents since it replaces the nonparametric DR (visualization component) with the parametric neural layers. 
In this paper, we validated the superiority of NeuralSI in inference speed over DeepSI, which is an essential property for interactive sensemaking tasks.
In addition, the neural design framework powers SI systems with other desired properties for visual analytics, including out-of-sample extensions and stability, as described in section~\ref{sec:mdp}.
Further explorations could be performed to balance both inference accuracy and computational efficiency of SI systems with interactive DL. 
One possible solution is to build a hybrid SI system with both DeepSI and NeuralSI to combine both advantages --- using DeepSI in the early stages for better inference accuracy (with a small number of data explorations) and NeuralSI in later stages for time efficiency for large data exploration. 



\subsection{Additional Neural Design Options for Semantic Interaction}
Further evaluation of the random initialization method in Section~\ref{sec:simulation} indicates that there remains room for improvement in developing accurate NeuralSI systems by exploring the design options of the projection head. 
For example, random initialization is a more appropriate parameter setting method for domain-specific tasks.
The new initialization method enables the NeuralSI system to own comparable inference ability as compared to DeepSI in $T_{\text{vis}}$.
Three design aspects we could explore to boost the NeuralSI framework performance include projection head architecture (linear vs. nonlinear projection head), projection parameter initialization (random vs. MDS-based initialization), and loss function (contrastive loss vs. MSE loss~\cite{espadoto2020deep,10.1145/3311790.3396646}).
Systematic studies of the effects of different design options and design combinations could be explored to improve the NeuralSI performance considerably.

\begin{figure}[htpb]
    \centering
  \includegraphics[width=0.85\textwidth]{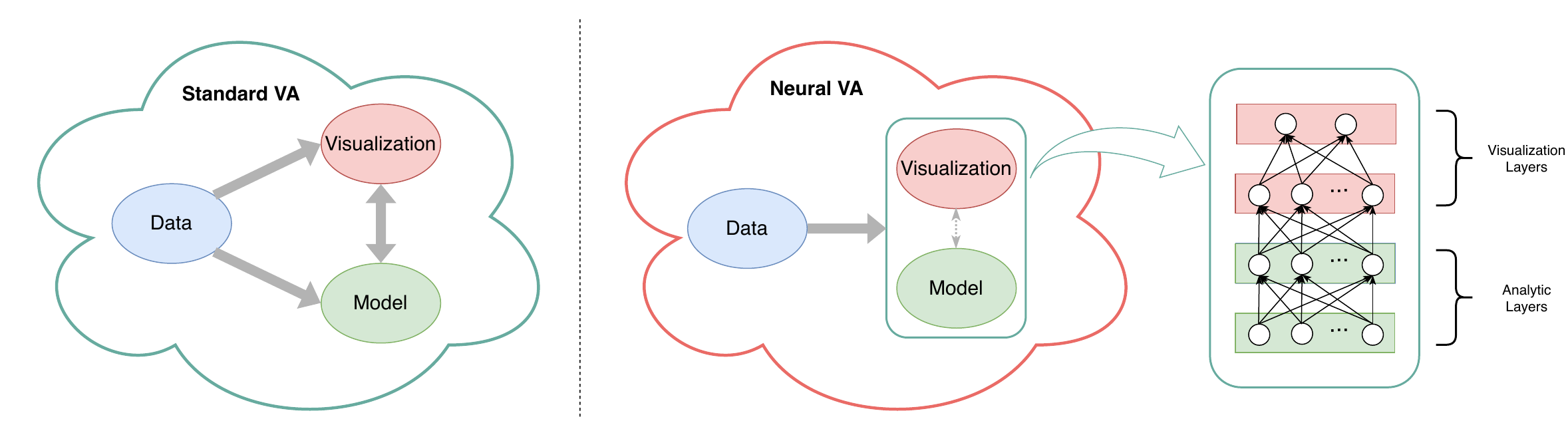}
  \caption{Comparison between the standard visual analytic framework and our proposed neural design framework of visual analytics for interactive deep learning. Standard visual analytics framework contains three separate components: data, visualization, and model. In contrast, in the neural design framework of visual analytics, the visualization and model components are merged into one integrated end-to-end neural network. Interactions applied to the visualization are directly passed to the backbone model for tuning through backpropagation.}
  \label{fig:end2end}
\end{figure}

\subsection{Towards the Neural Design of Visual Analytics}
NeuralSI is the end-to-end neural design of VA systems with SI for sensemaking tasks.
In NeuralSI, the neural projection head, appended to the DL backbone, works as the visualization method for interactive dimensionality reduction. 
More complex neural structures can be utilized as the visualization layer for other VA applications. 
Research on ML4VIS~\cite{10.1109/tvcg.2021.3106142}, which uses machine learning models to design compelling visualizations automatically, has received more attention recently.
In particular, neural networks have been used to generate visualizations.
These expressive neural networks can be appended to the DL backbone for other VA tasks. 
For example, Data2Vis~\cite{10.1109/MCG.2019.2924636} uses Seq2seq LSTM~\cite{10.1162/neco.1997.9.8.1735} to generate a diverse set of visualizations in the declarative language Vega-Lite\cite{2017-vega-lite}.
The Seq2seq LSTM can be appended to the underlying DL backbone as task-specific layers to build a neural VA system with a complex visual interface.
Therefore, more neural designs of the visualization component can be explored and appended to the DL backbone so as to build integrated trainable end-to-end deep neural networks as VA systems for other analytics tasks.
As shown in Fig.~\ref{fig:end2end}, we hope the neural design empowers VA systems with good properties of neural networks, such as better learning capability and efficient parallel computing using GPUs~\cite{LeCun:2015dt}.

\subsection{Limitations and Future Work}
For simplicity of comparison, this paper was limited to four datasets for visual text analytic tasks.
More experiments are needed to apply NeuralSI to a variety of analytics tasks, for example, adapting CNN models as the backbone for visual concept analytics~\cite{bian2019deepva}.
Further, a user study is needed to measure NeuralSI from the human perspective.
Specifically, we wish to investigate how NeuralSI systems boost human performance in complex sensemaking tasks~\cite{self2018observation,bian2019evaluating}.

In the future, we also plan to explore more advanced design choices in three aspects of NeuralSI:
the projection head architecture, projection parameter initialization, and loss function. 
First, more complex neural structures can be utilized in designing the projection head~\cite{DBLP:journals/corr/abs-1903-05987}.
For example, task-specific multi-layer neural networks, such as graph embedding~\cite{4016549}, could be utilized to generate data projections.
In addition, advanced initialization methods could be used to improve the learning performance of NeuralSI further.
For example, weight imprinting~\cite{qi2018lowshot} combines the best properties of the neural classifier with embedding approaches for solving the low-shot learning problem~\cite{chen2019closer}.
It can be redesigned for solving interactive DR tasks in our NeuralSI framework.
Finally, a particular loss function of NeuralSI could be designed for semantic interactions.
One solution could be using multiple loss functions~\cite{10.1145/3311790.3396646}, such as combining both the contrastive loss and the MSE loss.

\section{Conclusion}
This work focused on the neural design of SI systems for interactive deep learning in a framework called NeuralSI.
NeuralSI replaces both the analytic model and the visualization components with one integrated end-to-end trainable deep neural network.
The whole deep neural network can be divided into two parts based on functionality: the pretrained backbone for analysis and concept learning; the projection head for interactive dimensionality reduction. 
This design gives NeuralSI systems the substantial advantages of neural networks, such as out-of-sample extensions, stability, and inference speed.
We performed two complementary experiments to measure the effectiveness of NeuralSI, including a case study of a real-world task relating to COVID-19 and a simulation-based quantitative evaluation method on three commonly used text corpora.
The results of these two experiments demonstrated that NeuralSI has only slightly lower inference accuracy but clearly better time efficiency compared to the state-of-the-art alternative.

\begin{acks}
This work was supported in part by NSF I/UCRC CNS-1822080 via the NSF Center for Space, High-performance, and Resilient Computing (SHREC). This material is based upon work supported by the National Science Foundation under Grant \# 2127309 to the Computing Research Association for the CIFellows 2021 Project.
\end{acks}

\bibliographystyle{ACM-Reference-Format}
\bibliography{sample-base}




\end{document}